\documentclass[prl,twocolumn,superscriptaddress,showpacs,epsf]{revtex4}

\usepackage{graphicx}
\usepackage{latexsym}
\usepackage{amsmath}
\usepackage{amssymb}
\usepackage{amsfonts}
\usepackage{color}
\begin{document}

\title{Magnetic dipole induced guided vortex motion}

\author{N. Verellen}
\affiliation{INPAC-Institute for Nanoscale Physics and Chemistry,
Nanoscale Superconductivity and Magnetism $\&$ Pulsed Fields
Group, K. U. Leuven Celestijnenlaan 200 D, B-3001 Leuven,
Belgium.}

\author{A. V. Silhanek}
\affiliation{INPAC-Institute for Nanoscale Physics and Chemistry,
Nanoscale Superconductivity and Magnetism $\&$ Pulsed Fields
Group, K. U. Leuven Celestijnenlaan 200 D, B-3001 Leuven,
Belgium.}

\author{V. Metlushko}
\affiliation{Department of Electrical and Computer Engineering,
University of Illinois, Chicago, IL 60607.}

\author{W. Gillijns}
\affiliation{INPAC-Institute for Nanoscale Physics and Chemistry,
Nanoscale Superconductivity and Magnetism $\&$ Pulsed Fields
Group, K. U. Leuven Celestijnenlaan 200 D, B-3001 Leuven,
Belgium.}

\author{F. Gozzini}
\affiliation{Department of Electrical and Computer Engineering,
University of Illinois, Chicago, IL 60607.}

\author{B. Ilic}
\affiliation{Cornell Nanofabrication Facility, School of Applied
and Engineering Physics, Cornell University, Ithaca, New York
14853}

\author{V. V. Moshchalkov}
\affiliation{INPAC-Institute for Nanoscale Physics and Chemistry,
Nanoscale Superconductivity and Magnetism $\&$ Pulsed Fields
Group, K. U. Leuven Celestijnenlaan 200 D, B-3001 Leuven,
Belgium.}

\date{\today}
\begin{abstract}
We present evidence of magnetically controlled guided vortex
motion in a hybrid superconductor/ferromagnet nanosystem
consisting of an Al film on top of a square array of permalloy square
rings. When the rings are magnetized with an in-plane
external field ${\bf H}$, an array of point-like dipoles with
moments antiparallel to ${\bf H}$, is formed. The resulting
magnetic template generates a strongly anisotropic pinning
potential landscape for vortices in the superconducting layer.
Transport measurements show that this anisotropy is able to
confine the flux motion along the high symmetry axes of the square
lattice of dipoles. This guided vortex motion can be either
re-routed by 90 degrees by simply changing the dipole orientation or
even strongly suppressed by inducing a flux-closure magnetic state with very low
stray fields in the rings.
\end{abstract}

\pacs{74.78.-w 74.78.Fk 74.25.Dw}

\maketitle

During the last years there has been a considerable effort to
conceive and realize new superconducting devices that allow to
modulate locally the magnetic fields practically at will
\cite{zhu-2003}. These systems rely essentially on either static
arrays, such as pinning centers \cite{baert,fiory,martin,raedts}
and vortex-antivortex generators \cite{lange,gillijns}, or
components influencing the vortex dynamics like channels
\cite{martinoli,pastoriza,silhanek-03,villegas-06} and ratchets
\cite{villegas,clecio}. The ultimate motivation behind the
manipulation of the local vortex density is to enhance the
performance of superconductor-based devices by reducing the noise
in squid-based systems \cite{roger-02,selders}, gaining control on
superconducting THz emitters \cite{thz} or even providing a way to
predefine the optical transmission through the system
\cite{cond-mat}.

Unfortunately, for the majority of the components used in
fluxonics devices, once they are created there is no margin for
further modifications. In some cases this lack of flexibility
becomes a limiting factor in the performance of the devices. For
instance, a predefined ratchet system designed to effectively
remove vortices from a specific location can work properly at low
fields but ignores the inevitable reversed ratchet at higher
fields thus making its functionality rather impractical
\cite{villegas,clecio,lu,silhanekapl}. A way to circumvent this
shortcoming can be worked out by introducing magnetic pinning
centers which have additional internal degrees of freedom not
available in conventional nanostructured pinning sites or defects
created via irradiation.

In this work we demonstrate that reversible and switchable
guidance of the vortex motion can be achieved using a square
array of magnetic square rings lying underneath a superconducting
film. Transport measurements unambiguously show that the vortex
dynamics is fully dominated by the magnetic landscape generated by
the ring structures. When the magnetic unit cell is rotated 45
degrees off the Lorentz force ${\bf F}_L$ the average vortex
velocity ${\bf v}$ follows the direction of the principal axis of
the magnetic lattice rather than the driving force. This
channeling effect can be easily suppressed when inducing a nearly
isotropic pinning landscape by setting the square rings in a
flux-closure state with very low stray fields.

\begin{figure}[htb]
\begin{center}
\includegraphics[width=0.45\textwidth]{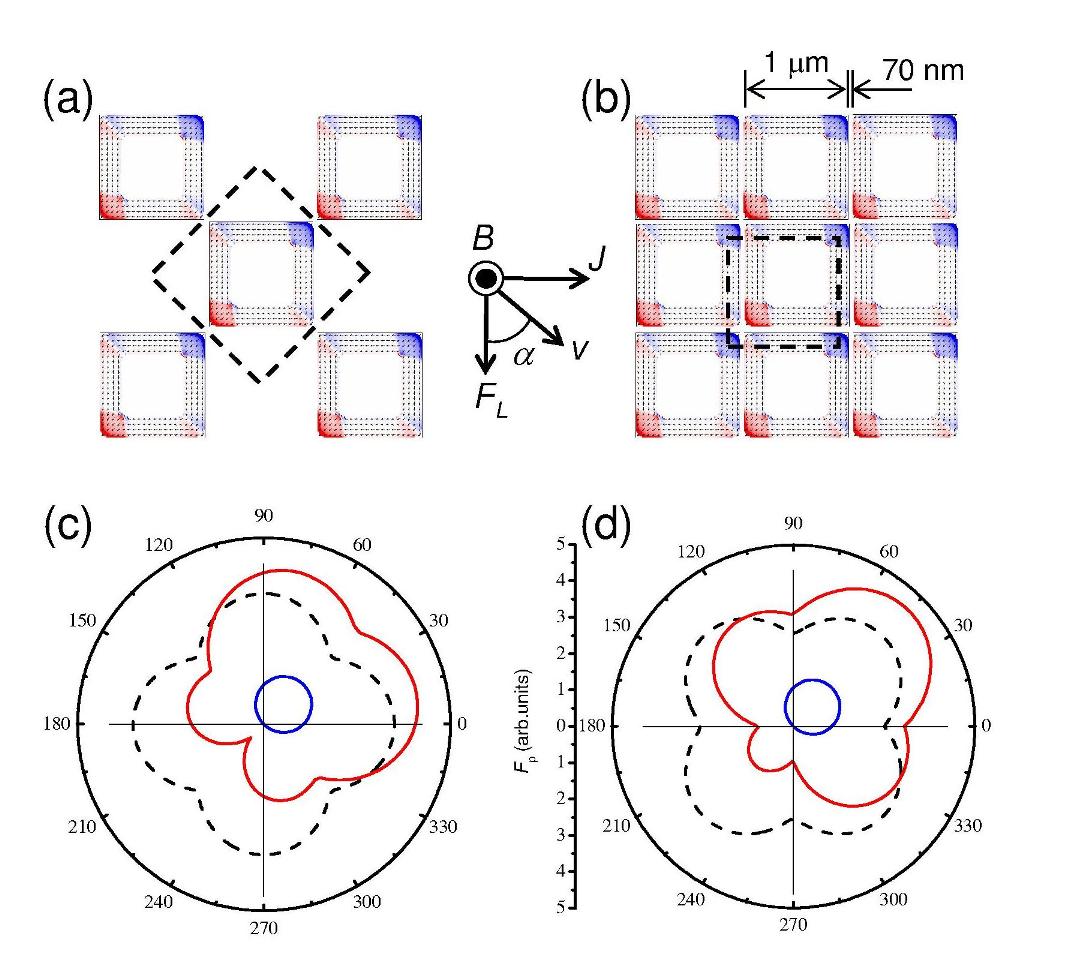}
\end{center}
\caption{(color on-line) Micromagnetic simulations of the out of
plane component of the stray field at remanence for the
checkerboard (a) and close-packed (b) arrays of square rings after
applying a saturation field along their diagonal (onion state).
The thick broken line indicates the magnetic unit cell for each
pattern. The orientation of the external current ($J$), field
($B$), driving force ($F_L$), and vortex drift ($v$) is clearly
indicated. The lower panels schematically show a polar plot of the
pinning force $F_p$ only considering the lattice symmetry (broken
line), or the local magnetic pinning force (off centered circle),
and including both the lattice and the local symmetry (solid line)
for the checker-board (c) and the close-packed (d) arrays.}
\label{fig1}
\end{figure}

The two samples used for this investigation consist of a 50 nm
thick Al film evaporated on top of a checker-board ($CB$)
patterned array and a close-packed ($CP$) square array of
permalloy square rings with lateral size 1 $\mu$m, line width 150
nm and thickness 25 nm [Fig. \ref{fig1}(a) and (b)]. The magnetic
template is electrically separated from the superconducting film
by a 5 nm Si buffer layer added to reduce proximity effects. The
ring-shaped patterns and the transport bridge were fabricated with
electron-beam lithography and lift-off technique on a silicon
substrate. A superconducting coherence length $\xi(0) \approx 130$
nm for the $CB$ sample and $\xi(0) \approx 138$ nm for the $CP$
sample was estimated from the superconducting/normal phase
boundary as determined by 10$\%$ normal state resistance criterion
of a coevaporated reference film. The superconducting critical
temperature at zero field for the CB and CP sample in the onion
state (i.e. maximum stray field) is $T_c = 1.356$ K and $T_c =
1.274$ K, respectively.

It has been shown recently that multiply connected magnetic structures make it possible to readily switch between
different magnetic states \cite{imre,Xiaobin-03}. For the
particular square geometry chosen in this work, an in-plane
external field along the diagonal of the squares can induce six
different domain distributions \cite{Vavassori-2003}, namely four
dipolar states (onion states) obtained at remanence after
saturation and two flux-closure states with opposite chirality
when the field is reduced from saturation to $\pm$ 36 mT and then
set to zero. Interestingly, when these rings are placed in close
proximity to a superconducting layer, either a strong vortex
pinning or weak pinning can be obtained by simply switching from
onion to flux-closure state, respectively. Although the influence
of these magnetic templates on the vortex pinning is relatively
well understood \cite{silhanek-06}, little is known about their
influence on the vortex dynamics.

In order to address this issue we carried out transport
measurements recording simultaneously the electric field parallel
($E_{xx}$) and perpendicular ($E_{xy}$) to the external current.
This allows us to estimate the direction $\alpha$ (see
Fig.\ref{fig1}) of the average vortex motion ${\bf v}$ with
respect to the Lorentz force ${\bf F}_L$. Notice that in contrast
to previous works studying the in-plane angular dependence of the
vortex motion \cite{pastoriza,silhanek-03,villegas-06} where the
orientation of the driving force was changed by combining two
independent current sources, our ability to change the potential
landscape \emph{in-situ} allows us to use a simpler setup where
the current orientation is kept constant. In all cases unwanted
contributions to the measured signal like thermoelectric coupling,
Hall resistance and misalignment of the transverse voltage
contacts are small and have been properly taken into account.

\begin{figure*}[!]
\begin{center}
\includegraphics[width=1\textwidth]{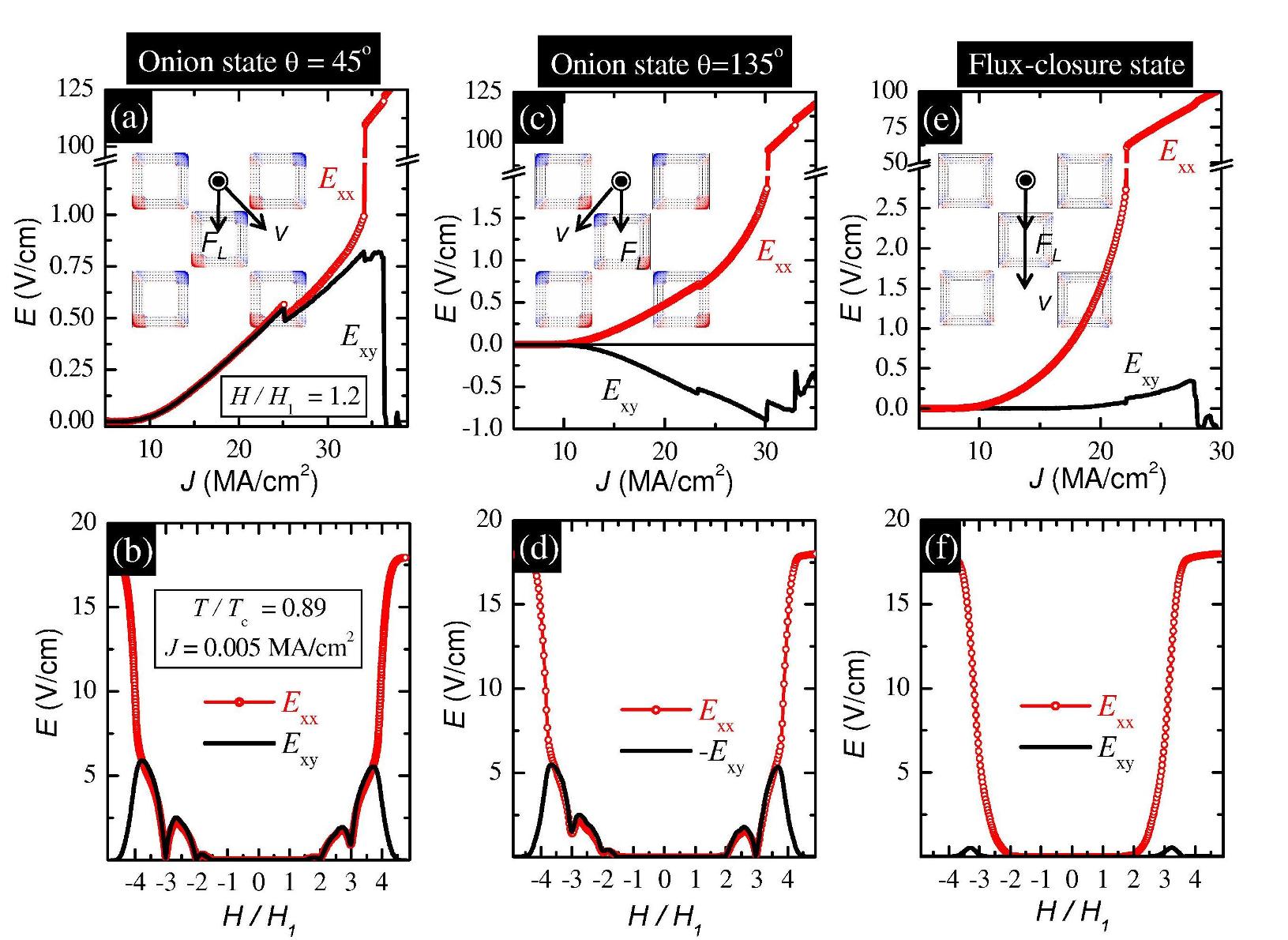}
\end{center}
\caption{(color on-line) Parallel ($E_{xx}$) and transverse
($E_{xy}$) electric field as a function of current density $J$
(upper row) and field (lower row) at $T/T_c=0.89$. The first
column [panels (a) and (b)] corresponds to the square rings
magnetized at 45$^\circ$ as indicated in the inset of panel (a).
The second column [panels (c) and (d)] presents the data for the
square rings magnetized at 135$^\circ$, as indicated in the inset
of panel (c). The third column [panels (e) and (f)] corresponds to
the squares in the vortex state, as indicated in the inset of
panel (e).} \label{fig2}
\end{figure*}

A representative set of experimental data for the $CB$ sample
[Fig.\ref{fig1}(a)] and different magnetic states of the square
rings is shown in Fig.\ref{fig2} at $T/T_c=0.89$.
Fig.\ref{fig2}(a) shows the electric field-current density ($E-J$)
characteristic for the case when the squares have been set in the
onion state magnetized 45$^{\circ}$ away from the current
direction [see inset in panel (a)] and setting an out-of-plane
field $H/H_1=1.2$ where $H_1=\Phi_o/d^2$, with $\Phi_o$ the flux
quantum and $d$ the period of the lattice. From Fig.\ref{fig2}(a)
it can be seen that for low enough currents ($J<8$ MA/cm$^2$) the
vortex lattice remains pinned as no dissipation is detected in
either direction ($E_{xx} = E_{xy} \sim 0$). Surprisingly, for
currents higher than the critical current the direction of vortex
motion does not coincide with the Lorentz force. More specifically
since $E_{xx} \approx E_{xy}$ vortices move at an angle
$\alpha=$45$^{\circ}$ as indicated in the inset of
Fig.\ref{fig2}(a). This is a clear evidence that the net
displacement of the vortices is along the high symmetry axis of
the magnetic pinning landscape indicated with a dashed line in
Fig.\ref{fig1}(a) and (b). Essentially this effect is the result
of an anisotropic depinning force $F_{dp}$ which reaches its
maximum $F_{dp}^{\perp}$ (minimum $F_{dp}^{||}$) value
perpendicular (parallel) to the channel direction defined by the
principal axes of the magnetic landscape. In Fig.\ref{fig2}(a),
since the applied force is in this case 45$^{\circ}$ away from the
channel direction, the critical current $J_{c1}$ at which vortices
start moving is determined by the condition $J_{c1} =
F_{dp}^{||}sin(45^\circ)/\Phi_o$. This situation persists up to
$J_{c2}\sim 25$ MA/cm$^2$ where $E_{xy}$ and $E_{xx}$ gradually
separate from each other which means that now the component of the
Lorentz force perpendicular to the channels is large enough to
overcome $F_{dp}^{\perp}$, i.e. $J_{c2} =
F_{dp}^{\perp}cos(45^\circ)/\Phi_o$, and vortices flow along the
direction of the Lorentz force. This allows us to estimate the
ratio of depinning forces as
$F_{dp}^{\perp}/F_{dp}^{||}=J_{c2}/J_{c1}\sim3$.

By keeping a constant current density and progressively increasing
the applied field it is possible to realize the same
dynamic behaviors, as shown in Fig.\ref{fig2}(b). For the low
current density $J = 0.005$ MA/cm$^2$ vortices
remain pinned up to a field $H/H_1 \sim 1.5$ where the
vortex-vortex interaction becomes stronger than the vortex-pinning
interaction. At higher fields clear commensurability effects can
be seen for $H/H_1 = 2$ and $H/H_1 = 3$. In this field region
($H/H_1 < 4$) the fact that $E_{xx} \approx E_{xy}$ indicates that
vortices are guided by the pinning potential as described above.

These findings clearly demonstrate that under certain conditions
the vortex motion can be effectively guided along the principal
axes of the underlying pinning potential, but tell us little about
the origin of this pinning potential. In other words, it is
necessary to find out whether the guided vortex motion (GVM) is
resulting from the magnetic landscape or from the corrugation of
the superconducting film deposited on top of the squares. In order
to answer this question we magnetized the square rings with an
in-plane field 135$^\circ$ away from the current direction
(0$^\circ$) as shown in the inset of Fig.\ref{fig2}(c). \emph{In
this way the topographic pinning remains the same whereas the
magnetic pinning is rotated 90$^\circ$ with respect to the
previous one.} If the magnetic landscape is responsible for the
GVM then a channeling along the -45$^\circ$ direction with respect
to ${\bf F}_L$ should be detected. This is in agreement with the
results shown in Fig.\ref{fig2}(c) and (d), where $E_{xx}$ remains
unchanged while $E_{xy}$ reverses sign. The most convincing
evidence however comes from panels (e) and (f) corresponding to
the square rings in the flux-closure state. In this case the stray
field generated by the domain walls in each square ring is reduced
to a minimum level and as a consequence it is expected a minor
influence from the magnetic landscape on the vortex dynamics. This
is indeed corroborated by the lack of guided motion as is
evidenced by the condition $E_{xx} >> E_{xy}$ \cite{comment}.

\begin{figure}[htb]
\begin{center}
\includegraphics[width=0.4\textwidth]{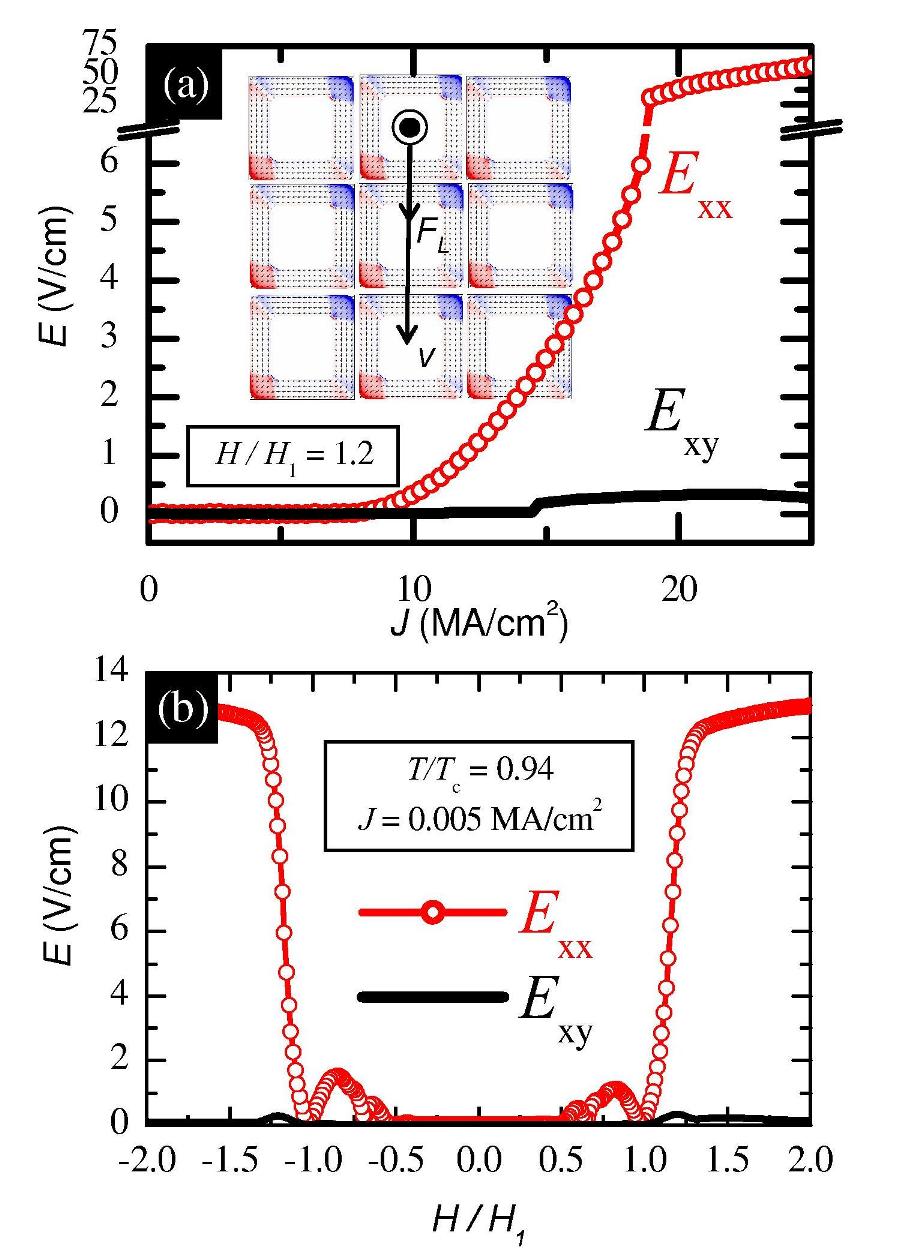}
\end{center}
\caption{(color on-line) Parallel ($E_{xx}$) and transverse
($E_{xy}$) electric field as a function of current density $J$
(upper panel) and field (lower panel) at $T/T_c=0.94$ for square
rings in a close-packed array and magnetized at 45$^\circ$ as
indicated in the inset.} \label{fig3}
\end{figure}

It is worth to note that the fact that for the same geometry of
the pinning lattice the GVM reverses sign when switching the
dipole orientation from $45^\circ$ to $135^\circ$, clearly points
out the relevance of the local symmetry of the pinning centra.
However, the question arises whether the geometry of the pinning
lattice plays any role in the vortex guidance. In order to tackle
this issue we have changed the lattice geometry from the checker-board
pattern to a close-packed array [see Fig.\ref{fig1}(b)]. In this
way we maintain a square lattice of nearly point-like dipoles
oriented at $45^\circ$ away from the current direction but change
the lattice period from 1.5 $\mu$m to 1.07 $\mu$m. In addition,
the unit cell of the dipolar lattice is rotated by $45^\circ$ as
illustrated in Fig.\ref{fig1}(a) and (b) with dashed lines.

The $E-J$ characteristics and the field dependence of the electric
field for this sample with the square rings in the onion state are
shown in Fig.~\ref{fig3} for similar experimental conditions shown
in Fig.~\ref{fig2}. Strikingly, the resulting measurements
indicate the lack of vortex guidance (i.e. $E_{xx} >> E_{xy}$) at
all temperatures and fields explored experimentally. This clearly
demonstrate the relevance of the particular orientation of the
magnetic unit cell for the GVM.

In order to understand the origin of the guided vortex motion lets
first hypothetically separate the different contributions arising
from the lattice symmetry and from the local symmetry of the
dipoles. This is schematically illustrated in panel (c) and (d) of
Fig.\ref{fig1} for the checker-board and close-packed arrays,
respectively. In order to unveil the lattice symmetry contribution
we can assume that each dipole behaves as an effective isotropic
pinning center, very much like a lattice of circular antidots or
magnetic dots. In this case it has been previously shown
\cite{silhanek-03} that the minimum pinning force is along the
principal axes of the pinning lattice [broken line in
Fig.\ref{fig1}(c) and (d)]. On the other hand, ignoring the
presence of the lattice and just considering the pinning force
produced by one single dipole gives rise to a strongly anisotropic
pinning force \cite{carneiro} [schematically displayed as an
off-centered circle in Fig.\ref{fig1}(c) and (d)]. It is precisely
the lack of inversion symmetry of the dipolar stray field which
makes these structures efficient ratchet systems as well
\cite{carneiro, clecio-2007}. The solid line in Fig.\ref{fig1}(c)
and (d) shows the resulting pinning force combining both effects:
the four-fold symmetric lattice contribution and the non-symmetric
local magnetic field. It is clear from this analysis that the
presence of the non symmetric magnetic landscape lifts the
degeneracy expected in the channeling of the $CB$ sample [see
Fig.\ref{fig1}(c)] as experimentally observed. Moreover, the lack
of guided motion in the $CP$ sample for the used current direction
is also expected since the minimum of the pinning force is along
the direction of the Lorentz force [see Fig.\ref{fig1}(d)],
irrespective of the asymmetries introduced by the magnetic
dipoles. It should be emphasized that although the above
description helps to identify the basic mechanisms behind the
guided motion of vortices, it is derived from the unphysical
assumption that magnetic and non-magnetic contributions are
separable.

In summary, we have reported experimental evidence indicating that
the anisotropic potential landscape of a square array of magnetic
dipoles is capable of directing the vortex motion 45 degrees away
from the driving force direction. The multiple states of the used
magnetic rings allow us to switch this direction ($\pm$ $45^\circ$
for two different dipolar states and $0^\circ$ for the
flux-closure state) resulting in a control of the transverse
voltage signal. The contrasting behavior of the two different
lattices explored (checker-board and close-packed arrays)
demonstrate that the guidance of vortices is not only a
consequence of the dipole orientation relative to the driving
force direction but that also the orientation of the square
lattice unit cell is important.

This work was supported by the Fund for Scientific
Research-Flanders FWO-Vlaanderen, the Belgian Inter-University
Attraction Poles IAP, the Research Fund K.U. Leuven GOA/2004/02,
the European ESF NES programs, the U.S. NSF, grant
ECS-0202780 and CNM ANL grants Nr.468
and Nr.470. A.V.S. is grateful for the support from the
FWO-Vlaanderen. We acknowledge useful discussions with C.C. Souza Silva and G. Carneiro.


\begin{references}
\bibitem{comment}
The small but finite transverse field observed when the rings are
set in the vortex state results from the fact that not all the
rings switch to that state but some (vast minority) remain in the
polarized state and therefore are still able to guide the vortices
but with much less efficiency.
\end{references}
\end{document}